\documentclass[
amsmath, amssymb, 
aps, pra,
reprint, twocolumn,
superscriptaddress,
longbibliography,
floatfix,
nofootinbib
]{revtex4}

\usepackage[a4paper, left=0.8in, right=0.8in, top=0.9in, bottom=0.9in]{geometry}

\usepackage[utf8]{inputenc}
\usepackage{amsmath,amssymb,amsthm}
\usepackage{braket}
\usepackage[urlcolor=blue, colorlinks=true,citecolor=blue,linkcolor=black,pagebackref=true]{hyperref}
\usepackage{cleveref}
\usepackage{graphicx}
\usepackage[caption=false]{subfig}
\usepackage{float}
\usepackage{tabularx}
\usepackage{amsfonts}
\usepackage{xcolor}
\usepackage{algorithm}
\usepackage{algpseudocode}
\usepackage{xy}
\usepackage{qcircuit}
\usepackage{tikz}
\usepackage[short]{optidef}

\renewcommand*\backref[1]{\ifx#1\relax \else (Cited on #1) \fi}

\begin{document}

\title{Parallel circuit implementation of variational quantum algorithms}
    
\author{Michele Cattelan}
    \affiliation{Volkswagen Data:Lab, Ungererstra\ss e 69, 80805 Munich, Germany}
    \affiliation{Institute for Theoretical Physics, University of Innsbruck, Innsbruck, A-6020, Austria.}
    
    \author{Sheir Yarkoni}
    \affiliation{Volkswagen Data:Lab, Ungererstra\ss e 69, 80805 Munich, Germany}

\begin{abstract}
    We present a method to split quantum circuits of variational quantum algorithms (VQAs) to allow for parallel training and execution, that maximally exploits the limited number of qubits in hardware to solve large problem instances. We apply this specifically to combinatorial optimization problems, where inherent structures from the problem can be identified, thus directly informing how to create these parallelized quantum circuits, which we call slices. We test our method by creating a parallelized version of the Quantum Approximate Optimization Algorithm, which we call pQAOA, and explain how our methods apply to other quantum algorithms like the Variational Quantum Eigensolver and quantum annealing. We show that not only can our method address larger problems, but that it is also possible to run full VQA models while training parameters using only one slice. These results show that the loss of information induced by splitting does not necessarily affect the training of parameters in quantum circuits for optimization. This implies that combinatorial optimization problems are encoded with redundant information in quantum circuits of current VQAs. Therefore, to attain quantum advantage for combinatorial optimization, future quantum algorithms should be designed to incorporate information that is free of such redundancies.
\end{abstract}

\maketitle

\section{Introduction}\label{sec: intro}
In recent years the field of quantum computing has grown in terms of interest due to its promise of being able to solve problems that are difficult or impossible to solve with classical computers~\cite{childs2003exponential, grover1996fast}.

Unlike classical computers, quantum computers work with a physical implementation that allows them to exploit some properties of quantum mechanics to create algorithms with improved complexity costs or efficiency. Various theoretical studies demonstrate that such an improvement can be reached and in some cases that a classical algorithm with the same performance cannot be developed~\cite{shor1999polynomial, childs2003exponential, grover1996fast, simon1997power}.


The current state of quantum hardware is not advanced enough to produce a computer able to make computations with an acceptable error rate. This is primarily due to the presence of noise, which causes errors during the computation on these machines. Consequently, over the past few years, various algorithms have been developed, known as variational quantum algorithms (VQAs), specifically designed to operate on such imperfect quantum machines~\cite{cerezo2021variational}. These algorithms fall under the class of hybrid quantum-classical algorithms, where a quantum circuit is implemented as a black-box function optimized using a classical method. Typically, circuits are characterized by a set of continuous real parameters in one and two-qubit gates whose values are determined by a classical optimizer. The goal is to find a set of parameters such that the output of the parameterized quantum circuit minimizes a given objective function. These algorithms present shallow and small circuits that, due to the reduced number of operations, are in general more resilient against noise~\cite{cerezo2021variational}.

Despite the challenges presented by the current technology, efforts are being made to make quantum computing practical and applicable to industrial problems. One promising application of quantum computing is to solve hard optimization problems and the sub-field devoted to this is called quantum optimization. Several algorithms were developed in the past with the aim of solving industrially relevant optimization problems~\cite{yarkoni2022quantum, zhou2020quantum}. 

The state of the art of quantum optimization algorithms requires the input problem to be encoded in a specific format, the Quadratic Unconstrained Binary Optimization (QUBO) problem. A QUBO involves finding a binary vector $x\in\mathbb{B}^n$ such that the value $x^TQx$ is minimized, where $Q$ is a symmetric matrix. This problem is equivalent to the Ising model, which is another physical problem that can be interchanged with QUBOs as input of a quantum algorithm, due to their equivalence. In the Ising model, the variables are spins that can take values of either +1 or -1 and are encoded in a symmetric matrix similar to QUBOs. The equivalence between these two models is given by a change of basis. In various applications of quantum optimization, these models have been demonstrated to be appropriate for representing various combinatorial optimization problems~\cite{Lucas2014}.

One of the algorithms that uses the Ising Hamiltonian to construct the circuit is the Quantum Approximate Optimization Algorithm (QAOA) which is inspired by the adiabatic theorem~\cite{farhi2014quantum}. Here, alternating layers of parameterized mixing and problem Hamiltonians are used in order to approximate ground states of a given problem Ising Hamiltonian. The parameters of the quantum circuit are then optimized classically in an outer loop with respect to the problem Hamiltonian, and the QAOA circuit acts as a black-box sampler in the inner loop. There are well-known theoretical results for both finite-depth and infinite-depth circuits which show how QAOA  can be used to solve some well-known optimization problems~\cite{zhou2020quantum}. Additional variants of QAOA have been developed in recent years to improve the performance of the algorithm under different optimization conditions (such as error mitigation, feasibility of solutions, etc.)~\cite{bravyi2020obstacles, bennett2021quantum, streif2020training, egger2021warm}.
A more general VQA approach to optimization is the Variational Quantum Eigensolver (VQE). Here, a quantum circuit is constructed given a problem-independent parameterized ansatz. Then, similarly to QAOA, the parameters of the ansatz are optimized with respect to a given Hamiltonian which represents some quantum system. In literature, VQE has been used to solve the minimum eigenvalue problem, which is equivalent to minimizing an Ising Hamiltonian representation of a combinatorial optimization problem, but can also be used to find ground states of other quantum systems~\cite{tilly2022variational}.

There are, however, significant limitations to the implementation of VQAs in state-of-the-art quantum hardware (also known as noisy intermediate-scale quantum processors, or NISQ~\cite{preskill2018quantum}) in the absence of error correction. The most relevant limiting factor is the number of qubits required to construct the circuits. Although minimizing an Ising Hamiltonian is NP-hard~\cite{barahona1982computational} and can be used to represent many combinatorial optimization problems of academic and practical interest, this often includes a polynomial overhead in the scaling of resources required to represent such problems~\cite{Lucas2014}. Typically, the limits of computability for hard problems are well beyond those that can be solved with existing quantum hardware, and so only small toy instances of said problems are typically solved with VQAs. Furthermore, due to high error rates and low coherence times, especially at high depths, the effect of noise becomes non-negligible and reduces the performance of VQAs~\cite{noiseVQA}. Lastly, it is important to note that the quality of the results of these VQAs depends on the optimization of the parameters in the quantum circuit by definition. Furthermore, it has been shown that training these parameters optimally is in itself an NP-hard problem~\cite{bittel2021training}, and as such, implying that finding optimal VQA parameters is at least as hard as solving the combinatorial optimization problems themselves.

In this work, we attempt to mitigate the limitations of VQAs by presenting a novel method to parallelize any variational quantum algorithm. For the sake of simplicity, we motivate our method by considering quantum optimization algorithms in particular, although our method is generalizable to all variational quantum algorithms. The essence of the method is that we approximate the output state of the tensor product of the unitary matrices composing the quantum circuit into a Cartesian product of output states from smaller quantum circuits. In other words, given a combinatorial optimization problem and a VQA ansatz, we approximate the ground state distribution of the problem by splitting the ansatz into independent smaller parameterized quantum circuits, each of which is optimized in parallel and guided by a single global objective function. The result of this procedure is a collection of classically separable quantum systems with shallow circuits whose product of vector spaces matches the original optimization search space of the problem.

The paper is structured as follows: in \cref{sec: related work} we discuss previous results that motivate our method; in \cref{sec: parallelize} we demonstrate the derivation of our method explicitly using QAOA as a starting point, and then show how to extend the method to other VQAs; in \cref{sec: VRP} we present one such constrained combinatorial optimization problem, the vehicle routing problem, that we use as an example because of symmetric properties of its QUBO formulation; in \cref{sec: numerical results} we test our parallelized version of QAOA experimentally providing both insights into the physical significance of our parallelization technique as well as benchmark its performance with respect to well-known combinatorial optimization techniques.

\section{Related works}\label{sec: related work}
One of the questions of practical relevance that quantum optimization researchers are trying to solve is how to efficiently implement constrained optimization problems in quantum computers. In particular implementing constraints requires an overhead of resources in terms of qubits and interactions between them. Therefore, the implementation of larger problems becomes impractical because the number of qubits and the implementable circuit size are insufficient to meet the requirements of such problems. Hence, while encoding a constraint, we have to minimize the number of additional quantum resources required to implement it. Along this direction in~\cite{hao2022exploiting, wang2022quantum} possible solutions are presented. The authors proposed to not implement the constraint as part of the Hamiltonian that defines the circuit, but rather implement it as part of the function used to optimize the hyperparameters. This results in transferring the information regarding the feasibility of eh constraints from the quantum simulation to the classical search showing improvement in both the quality of the solutions and in overlap with the solution state. Additional work in this direction using VQE employs the concept of \emph{contextual subspaces} in molecular simulations, where a quantum Hamiltonian is split into two separable Hamiltonians, whose sum reconstructs the original Hamiltonian of interest~\cite{Kirby2021}. One of these Hamiltonians is then computed classically, and the second attempts to ``correct'' the classical approximation using a VQE method. While the authors note that the classical simulation component is still NP-hard, the number of qubits required to implement such a hybrid method was significantly smaller compared to other VQE methods, while still maintaining the chemical accuracy of the model.

Another viable approach to handle the overhead of qubits and gates while implementing constraints is to apply circuit cutting and knitting techniques~\cite{wang2022quantum,tuysuz2022classical}. Although these methods show that we can simulate larger quantum systems using fewer qubits and achieve improved solution quality in certain scenarios, this outcome involves a trade-off. Indeed, in both cases the authors highlight that there is an exponential overhead in the number of measurements we have to apply in order to reconstruct the correct wavefunction~\cite{wang2022quantum} or in the number of cuts we can apply to the circuit, resulting in an exponential search to find the best and suitable way of cutting the circuit~\cite{tuysuz2022classical}.

\section{Parallelizing variational quantum algorithms for optimization}\label{sec: parallelize}
In this section, we propose a method to create a parallelizable algorithm from a VQA. The procedure is inspired by some of the results presented in \cref{sec: related work}.
We can formally define VQAs as a parameterized quantum circuit to optimize over the pseudo boolean classical function $H:\{0, 1\}^{n}\rightarrow\mathbb{R}$. 
The circuit is initialized with a set of parameters and the final state is sampled. The samples are used to evaluate $H$ and compute its gradient. These results are used by a classical optimizer to update the parameters of the quantum circuit until convergence or a desired result is reached.
To evaluate the pseudo boolean function $H$, we consider the quantum states measured in the computational basis as bit-strings.
In fact, the quantum subroutine is designed to have outcomes that belong to the same search space of $H$, resulting in a matching between the space where the circuit outcomes belong and the domain where $H$ is defined. 

Due to the scarce resources available in the state of the art of quantum hardware, implementing problems of a large size can often be impractical. Our proposed approach tackles this issue by creating several smaller quantum circuits that are tailored to the properties of the problem and that can be executed on the available resources. This means that, instead of having a one-to-one correspondence between the output of the single circuit and the function to optimize, we introduce a representation of the search space based on products of subspaces.


We now explain the general method to create a parallelization of quantum algorithms by inspecting the problem directly. We call this approach \emph{slicing}. We consider a VQA that is described by a quantum system of $N$ qubits and a quantum hardware that has $n$ qubits available, where we want to implement the quantum circuit of the VQA in. By inspecting the problem, we identify $k$ different subsystems, called \emph{slices}, of maximum dimension $n$, whose product matches the original output space of the VQA, which is the search space of our problem. Notice that the outer classical optimization routine is no longer optimizing a black box defined on a $2^{N}$-dimensional space, but $k$ black boxes defined on spaces of dimension at most $2^{n}$. 

Now, let us distinguish two different cases. If $N>n$, implementing the original circuit requires more qubits than the number available in the hardware, therefore the algorithm can only be implemented in its parallelized version. On the other hand, when $N\le n$ we can see that even though the circuit can now be implemented, our method reduces the number of interactions used. Therefore, in both cases, our approach presents a reduction in the number of resources used.

Given the above method, we can, specifically, formalize our quantum circuit as the following function 
$$\mathcal{C}:\mathbb{R}^{q}\rightarrow \mathbb{U}(2^n),$$
where $\mathbb{R}^q$ is the search space of the parameters and $\mathbb{U}(2^n)$ is the space of unitary matrices of dimension $2^n$. The function $\mathcal{C}$ fulfills the following property:
$$\mathcal{C}(\alpha_1,\ldots,\alpha_q)=U_1\otimes\cdots\otimes U_r,$$
where $U_1, \ldots, U_r\in \bigcup_{j=1}^{n}\mathbb{U}(2^j)$.

\subsection{A parallel Quantum Approximate Optimization Algorithm (pQAOA)}\label{subsec: pQAOA}
We now explain how to inspect a combinatorial optimization problem to create parallel slices for VQA by taking QAOA as an example. We stress that this procedure can be used for any VQAs. Consider the QAOA ansatz for finite depth $p$:
\begin{equation}\label{eq: QAOA ansatz}
    e^{-i\beta_p H_i}e^{-i\gamma_p H_f}\cdots e^{-i\beta_1 H_i}e^{-i\gamma_1 H_f}.
\end{equation}

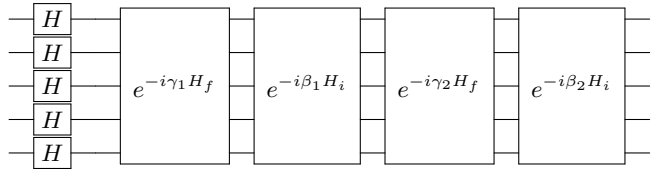
\begin{figure}
    \centering
    \begin{displaymath}
        \begin{array}{c}
             \Qcircuit @C=1em @R=0.1em {
                & \gate{H} & \qw & \multigate{4}{e^{-i\gamma_1 H_f}} & \multigate{4}{e^{-i\beta_1 H_i}} & \multigate{4}{e^{-i\gamma_2 H_f}} & \multigate{4}{e^{-i\beta_2 H_i}} & \qw \\
                & \gate{H} & \qw & \ghost{e^{-i\gamma_1 H_f}} & \ghost{e^{-i\beta_1 H_i}} & \ghost{e^{-i\gamma_2 H_f}} & \ghost{e^{-i\beta_2 H_i}} & \qw\\
                & \gate{H} & \qw & \ghost{e^{-i\gamma_1 H_f}} & \ghost{e^{-i\beta_1 H_i}} & \ghost{e^{-i\gamma_2 H_f}} & \ghost{e^{-i\beta_2 H_i}}  & \qw\\
                & \gate{H} & \qw & \ghost{e^{-i\gamma_1 H_f}} & \ghost{e^{-i\beta_1 H_i}} & \ghost{e^{-i\gamma_2 H_f}} & \ghost{e^{-i\beta_2 H_i}} & \qw\\
                & \gate{H} & \qw & \ghost{e^{-i\gamma_1 H_f}} & \ghost{e^{-i\beta_1 H_i}} & \ghost{e^{-i\gamma_2 H_f}} & \ghost{e^{-i\beta_2 H_i}}& \qw\\
            } 
        \end{array}
    \end{displaymath}
    \caption{Level-2 QAOA circuit. The initial state of the circuit is $|0\rangle^{\otimes_n}$. The circuit for larger $p$ is obtained by sequential repetition of the two layers as described in \cref{eq: QAOA ansatz}.}
    \label{fig: vanilla QAOA}
\end{figure}

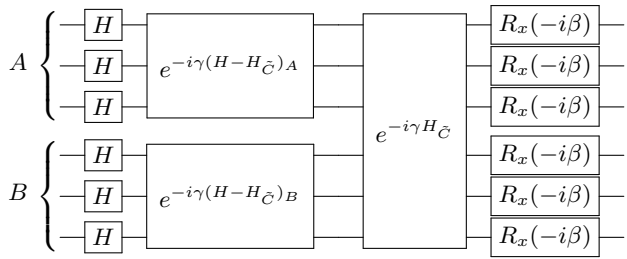
\begin{figure}
    \centering
    \begin{displaymath}
        \begin{array}{c}
             \Qcircuit @C=1em @R=0.1em {
                & \gate{H} & \multigate{2}{e^{-i \gamma (H-H_{\tilde{C}})_{A}}} & \qw & \multigate{8}{e^{-i \gamma H_{\tilde{C}}}} & \gate{R_x(-i \beta)} & \qw \\
                & \gate{H} & \ghost{e^{-i \gamma (H-H_{\tilde{C}})_{A}}} & \qw & \ghost{e^{-i \gamma H_{\tilde{C}}}} & \gate{R_x(-i \beta)} & \qw \\
                & \gate{H} & \ghost{e^{-i \gamma (H-H_{\tilde{C}})_{A}}} & \qw & \ghost{e^{-i \gamma H_{\tilde{C}}}} & \gate{R_x(-i \beta)} & \qw 
                    \inputgroupv{1}{3}{1em}{1.5em}{A} \\
                & & & & & \\
                & & & & & \\
                & & & & & \\
                & \gate{H} & \multigate{2}{e^{-i \gamma (H-H_{\tilde{C}})_{B}}} & \qw & \ghost{e^{-i \gamma H_{\tilde{C}}}} & \gate{R_x(-i \beta)} & \qw \\
                & \gate{H} & \ghost{e^{-i \gamma (H-H_{\tilde{C}})_{B}}} & \qw & \ghost{e^{-i \gamma H_{\tilde{C}}}} & \gate{R_x(-i \beta)} & \qw \\
                & \gate{H} & \ghost{e^{-i \gamma (H-H_{\tilde{C}})_{B}}} & \qw & \ghost{e^{-i \gamma H_{\tilde{C}}}} & \gate{R_x(-i \beta)} & \qw 
                    \inputgroupv{7}{9}{1em}{1.5em}{B}
            } 
        \end{array}
    \end{displaymath}
    \caption{The level-1 QAOA circuit implementation of \cref{eq: constrained hamiltonian}. Notice that the Hamiltonian $H-H_{\tilde{C}}$, and therefore its exponential, is separable between register $A$, first $n$ qubits, and register $B$, last $m$ qubits.}
    \label{fig: separable}
\end{figure}

The QAOA circuit with a generic Hamiltonian $H_f$ is as shown in \cref{fig: vanilla QAOA}. For simplicity, we start by considering constrained optimization problems, and then, in \cref{sec: non constraints}, explain how our method generalizes. Constrained combinatorial optimization problems can be represented by the following Hamiltonian:

\begin{equation}\label{eq: constrained hamiltonian}
    H=H_\mathrm{obj} + H_C,
\end{equation}
where $H_\mathrm{obj}$ represents the objective function of the combinatorial optimization problem that we are considering and $H_C$ is the Hamiltonian that encodes the constraints that define the feasible region of the problem. Let $N$ be the number of qubits in the QAOA circuit and let us consider the Hamiltonian $H_{\tilde{C}}$ defined on all $N$ qubits, that implements some constraints of the problem\footnote{Note that constraints in QUBOs are penalty terms add as addends to the objective function.}. Further, we assume that by removing $H_{\tilde{C}}$ we create two classically separable Hamiltonians that operate on two registers which we call $A$ and $B$, of length $n$ and $m$ respectively (such that $n+m=N$), see \cref{fig: separable}. Note that the circuits created in this way do not fully represent the original problem. Additionally, note that in our method it is sufficient only to include in $H_{\tilde{C}}$ the minimum set of constraints necessary in order to create these classically separable Hamiltonians. Therefore, we must modify the classical subroutine in order to implement the information missing from $H_{\tilde{C}}$. 
To solve this, we apply the same argument as in~\cite{hao2022exploiting}: we can leave $H_{\tilde{C}}$ out of the quantum circuit, implement the quantum circuit according to the Hamiltonian $H-H_{\tilde{C}}$ and execute the classical optimization subroutine by evaluating the ``complete" Hamiltonian function $H$, that can be trivially read as $\left(H-H_{\tilde{C}}\right) + H_{\tilde{C}}$. Notice that even though the information about the feasibility of the constraints encoded by $H_{\tilde{C}}$ is now only part of the classical optimizer, we are still seeking solutions that fulfill $H_{\tilde{C}}$ since we are minimizing $H$. 

This procedure results in two separate quantum circuits of size $n$ and $m$ that can be executed independently on separate quantum registers. Note that the two quantum circuits depend on parts of the Hamiltonian $H-H_{\tilde{C}}$ that do not share any terms and, hence, their circuit representations are separable. We call such \textit{classically separable} Hamiltonians and each of the circuits define by them a \textit{slice}. 

Note that the solutions from register $A$ are $n$-dimensional vectors, whereas the ones from $B$ are $m$-dimensional vectors. Let $S_A$ ($S_B$) be the multiset\footnote{A multiset is a set where elements can be repeated more than once.} of samples from register $A$ ($B$). To construct samples overall $N$ qubits using the $n$- and $m$-dimensional solutions, we take the product in the following way: 

\begin{multline}\label{eq: juxtapose}
|s\rangle\in S =\{|s_A\rangle\otimes |s_B\rangle : (|s_A\rangle, |s_B\rangle)\in S_A \times S_B\}.
\end{multline}

This mapping from slice samples to full Hamiltonian samples is sufficient to construct a parallel implementation for QAOA. Further note that, due to the fact that the slices are classically separable quantum circuits, we can measure them independently and therefore also the optimization of their respective parameters can be done independently. Therefore, we can choose whether to parameterize each slice independently or keep the same number of parameters as in the original QAOA ansatz. For the remainder of this discussion, we do not assume either case and simply refer to the parameters of the ansatzes as $\vec{\gamma}$ and $\vec{\beta}$-- our description holds for both.

In order to incorporate $H_{\tilde{C}}$ into the optimization of $\vec{\gamma}$ and $\vec{\beta}$, we must evaluate it classically. However, note that by removing $H_{\tilde{C}}$ we relax one of the assumptions of QAOA: we are no longer interested in ground states of the individual slices, but rather ground states of the original global Hamiltonian $H$, which can now be excited states of the slices. Therefore, in order to guide the optimization procedure of $\vec{\gamma}$ and $\vec{\beta}$ to the global optimum, we evaluate $H$ with our composed solutions $S$. Meaning, if $\psi$ is the wavefunction of the original QAOA circuit for $H$, then we approximate it by minimizing the following expectation value:

\begin{equation}
    {\langle{\psi}|H|{\psi}\rangle} \approx {\langle{S}|H|{S}\rangle},
\end{equation}
where $\ket{S}= \frac{1}{\sqrt{|S|}}\sum_{\ket{s}\in S} \ket{s}$.
The expectation value minimized in this way is now evaluated by the Hamiltonian $H$ and, thus, the parameters $\vec{\gamma}$ and $\vec{\beta}$ are updated with respect to the original optimization problem. The slices of the pQAOA algorithm now function as independent black boxes used to sample separable regions of the search space. 

\begin{figure}
    \centering
    \includegraphics[scale=0.2]{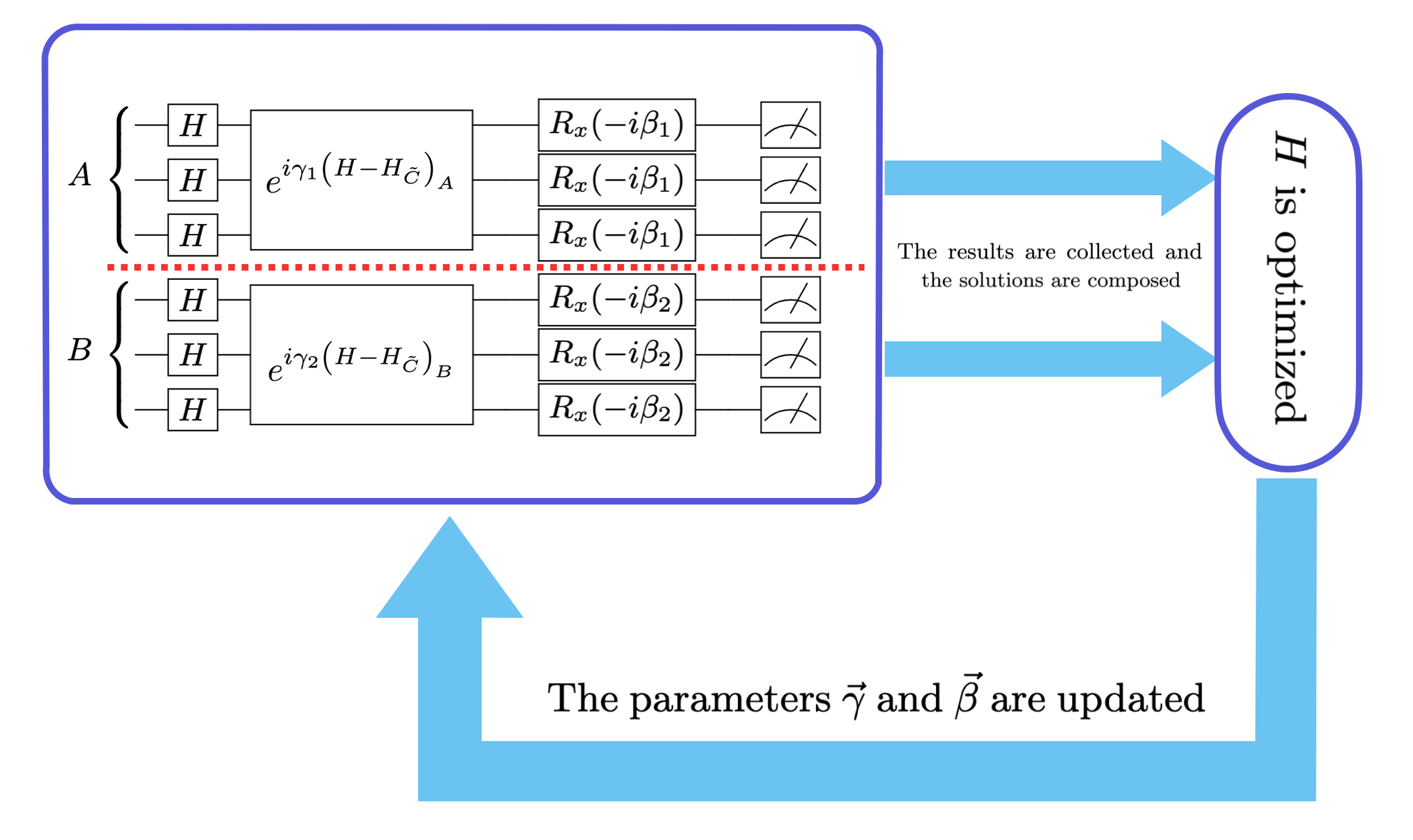}
    \caption{The pQAOA obtained by using our method on the Hamiltonian \cref{eq: constrained hamiltonian}. The two register $A$ and $B$ are not connected by any gates and they can be executed in parallel. The red dashed line stress this fact. This implies that we can use only $\max\{m, n\}$ qubits to execute this algorithm. The results are collected from the two slices and glued together. In this way, we obtain states that can be evaluated with $H$. The parameters are, then, optimized and the algorithm can proceed with the next iteration. In the figure, we use $\beta_1, \beta_2$ and $\gamma_1, \gamma_2$, but to be consistent with the QAOA parameterization one can also use only one $\beta$ and one $\gamma$.}
    \label{fig: pQAOA}
\end{figure}

The whole pQAOA is illustrated in \cref{fig: pQAOA} with $p=1$ as an example (although generalizing this procedure for any $p$ with additional layers is trivial).


\subsection{A parallelization for the variational quantum eigensolver}
The variational quantum eigensolver (VQE) is a hybrid quantum-classical variational quantum algorithm used to solve the minimum eigenvalue problem. For a given initial state and choice of parameterized ansatz, a quantum circuit is defined with the goal of iteratively adapting ansatz parameters to minimize a target objective function. Specifically, given a unitary representation of the ansatz, $U(\vec{\theta})$, where $|\vec{\theta}|$ is the number of parameters in the ansatz, VQE attempts to solve:
\begin{equation}
    \mathrm{argmin}_{\vec{\theta}}\bra{\psi(\vec{\theta})}\mathcal{H}\ket{\psi(\vec{\theta})},
\end{equation}
where $\psi(\vec{\theta})=U(\vec{\theta})\ket{0}$ is the ansatz applied to the initial state (in this case the zero state $\ket{0}$), and $\mathcal{H}$ is a Hamiltonian whose minimum eigenvalue we wish to find (or approximate). Originally proposed in~\cite{peruzzo2014variational}, VQEs have been used in practice to solve a variety of problems in quantum chemistry and combinatorial optimization~\cite{peruzzo2014variational,LayerVQE,FilterVQE}. However, similar to other VQAs, VQE suffers from the same limitations of qubit count, circuit depth, and parameter optimization which limit its usefulness in applications. To overcome this, using our parallelization technique, we propose a variant, pVQE, which we outline here. 

One of the strengths of VQE is the freedom in the choice of ansatz. The goal is to construct an ansatz that can simultaneously be expressive enough to explore the Hilbert space of the circuit as well as be easily implementable~\cite{NoiseVQE}. We note that, for our case of pVQE, we do not require the quantum Hamiltonian to explore the entire search space of the original problem, but only within each slice. Therefore, we have even more freedom with respect to the original VQE implementation.
Consider the hardware-efficient ansatz (HEA), a common choice of ansatz for VQE due to its ease of implementation~\cite{VQEHWE}. For $L$ layers of the VQE circuit with $N$ qubits, we have $\mathcal{O}(NL)$ parameters and $\mathcal{O}((N-1)L)$ entangling gates. However, for pVQE with $N = kn$ qubits and $L$ layers, we have only $\mathcal{O}(k(n-1)L)$ entangling gates. Furthermore, most critically, while we have the same total number of parameters $\mathcal{O}(NL) = \mathcal{O}(knL)$, each slice now occupies a Hilbert space of only $2^n$. Meaning, the pVQE HEA within each slice with $\mathcal{O}(nL)$ parameters needs to explore a space of $2^n$, compared to the original VQE which needs $\mathcal{O}(knL)$ parameters to explore a space of $2^{kn}$. While this example specifically exploits the HEA, it generally holds that the Hilbert space of pVQE is exponentially smaller than that of VQE, which pVQE can exploit more efficiently, thus alleviating the trade-off between expressivity and each implementation of ansatz.

\subsection{A parallelization for quantum annealing}
Quantum annealing (QA) is one of the original quantum optimization algorithms designed to solve combinatorial optimization problems by exploiting adiabatic evolution, both in simulation and in programmable quantum hardware~\cite{nishimori,Johnson2011}. The algorithm works by initializing a quantum system (or simulation) to an easy-to-prepare ground state (initial Hamiltonian $H_i$), and then evolving the system to represent a different Hamiltonian (final Hamiltonian $H_f$) to be minimized. The result is a metaheuristic optimization algorithm that can be used to simulate quantum Hamiltonians. 
In-depth technical works on the physics behind quantum annealing theory and its implementation in hardware can be found in:~\cite{Hauke2020}. For the purposes of implementing a parallelized version of QA, we only introduce the necessary components for constructing our algorithm, and encourage the interested reader to review the works cited above for more information.

The most commonly used Hamiltonian for quantum annealing is known as the transverse-field Ising Hamiltonian:
\begin{equation}
    \mathcal{H}(s) = A(s)\left[ \sum_{i}\sigma_i^x\right] + B(s)\left[ \sum_{i}h_i \sigma_i^z + \sum_{i<j} J_{ij} \sigma_i^z \sigma_j^z \right].
\end{equation}
Here, $s = t/\tau \in [0, 1]$ is referred to as ``normalized time'', and $\tau$ is the duration of evolution, an input parameter to the algorithm. The initial Hamiltonian shown here is $H_i = \sum_{i}\sigma_i^x$ and the final Hamiltonian $H_f$ is the quantum Ising Hamiltonian with z-spin Pauli operators $\sigma_i^z, \sigma_j^z$. The objective of the QA algorithm is to minimize $H_f$, which can be used to represent NP-complete and NP-hard problems and is therefore of practical interest~\cite{Lucas2014}. A review of previous work in applying quantum annealing in practice can be found here~\cite{yarkoni2022quantum}.

In this paper, we focus on motivating one method of constructing a parallel version of QA by exploiting a parameter known as \emph{annealing offsets}. This specific parameter allows for the advancement or delay of the point in (normalized) time at which each qubit in the quantum annealer starts its evolution from $H_i$ to $H_f$. This shift is denoted by $\Delta s_i$ for qubit $i$, with $\Delta s_i > 0$ being a delay, and $\Delta s_i < 0$ being an advancement. Due to the changing eigenspectrum generated by the evolving Hamiltonian, it is known that some qubits in the system experience a slowdown in tunneling dynamics before the termination of the evolution, an effect known as ``freeze-out''. This is known to affect the performance of QA as it makes it harder for the system to remain in the ground state~\cite{freezeout1,freezeout2}. The annealing offset parameter can therefore be used to change the freeze-out point on a per-qubit basis, in an attempt to mitigate this effect by synchronizing the freeze-out points. It has been demonstrated in quantum annealing hardware that tuning these parameters can (sometimes significantly) improve the probability of observing the ground state of $H_f$~\cite{LantingOffsets,integeroffsets,Yarkoni2019}. In general, however, given that the offsets are continuous parameters with non-convex search space, tuning these parameters optimally is a hard problem in itself.

To implement a parallel QA algorithm (pQA), we use a similar paradigm as for the previous algorithms presented above. We start by constructing the global and local Hamiltonians for our optimization problem. For each slice, we parameterize the annealing offsets for each qubit in the problem independently and tune them within each slice using the global Hamiltonian as the target function. 


While this proposal doesn't reduce the number of offset parameters in the problem (we use the same number of qubits for pQA as in QA), this does have a significant physical effect on the search space. Since each slice is embedded on the quantum annealer independently, we are only attempting to mitigate the freeze-out \textit{within} each slice. Therefore, the search space is much more confined with respect to the global Hamiltonian. 


\section{The vehicle routing problem}\label{sec: VRP}
We examine the Vehicle Routing Problem (VRP), a well-known NP-hard optimization problem, as the testbed for our parallelized VQAs. As described in Sec.~\ref{sec: parallelize}, we can exploit the constraints in the problem formulation in order to directly inform how to build the circuit slices in our quantum implementation. In the VRP, we consider a fleet of vehicles that need to deliver goods or services to a set of customers. The goal is to find the optimal set of routes for the vehicles that will minimize costs associated with the deliveries. The QUBO definition of the VRP is as follows~\cite{feld2019hybrid, marsh2019quantum, cattelan2022modeling}:

\begin{multline}\label{eq: VRP}
    H=H_{\mathrm{of}}+\sum_{i=1}^{n}\left(1-\sum_{a=0}^{A-1}\sum_{s=0}^{n} x_{a,i,s} \right)^{2}
    + \\ +\sum_{a=0}^{A-1}\sum_{s=0}^{n}\left(1-\sum_{i=0}^{n} x_{a,i,s} \right)^{2},
\end{multline}
with 
\begin{equation*}
    H_{\mathrm{of}}=\sum_{a=0}^{A-1}\sum_{i,j=0}^{n}\sum_{s=0}^{n}\frac{w_{i, j}}{W}x_{a,i,s}\,x_{a,j,s+1},
\end{equation*}
where the locations, $i$, are numbered from $0$ to $n$ and $0$ is the depot, i.e. the location where the vehicles start; $w_{i,j}$ are the costs associated to reach location $j$ from location $i$ and $W:=\max_{i,j} w_{i,j}$; $A$ is the number of vehicles;  and, the index $s$ represents the discrete step of the process.\footnote{This is the algebraic description of the QUBO, i.e. we write directly the polynomial $x^T Qx$.}

By considering the QUBO formulation of the problem in \cref{eq: VRP}, we can see that only the second addend contains quadratic terms that involve different indices for the vehicle, $a$. In addition, we stress the fact that this property yields symmetry in the problem that can be exploited to construct the slices. Indeed, we can write $H$ in the following fashion:
\begin{multline*}
    H=
    \sum_{a=0}^{A-1}\left[\sum_{i,j=0}^{n}\sum_{s=0}^{n}\frac{w_{v,k}}{W}x_{a,i,s}\,x_{a,j,s+1} + \right.\\ 
    + \left. \sum_{s=0}^{n}\left(1-\sum_{i=0}^{n} x_{a,i,s} \right)^{2} \right]+\\
    +\sum_{i=1}^{n}\left(1-\sum_{a=0}^{A-1}\sum_{s=0}^{n} x_{a,i,s} \right)^{2},
\end{multline*}
and by considering
\begin{align*}
    H_a &= \sum_{i,j=0}^{n}\sum_{s=0}^{n}\frac{w_{v,k}}{W}x_{a,i,s}\,x_{a,j,s+1}
    + \sum_{s=0}^{n}\left(1-\sum_{i=0}^{n} x_{a,i,s} \right)^{2},\\
    H_c &= \sum_{i=1}^{n}\left(1-\sum_{a=0}^{A-1}\sum_{s=0}^{n} x_{a,i,s} \right)^{2},
\end{align*}
we can summarize $H$ as:
\begin{equation*}
    H=H_c+\sum_{a=0}^{A-1}H_{a}.
\end{equation*}
One can notice that the Hamiltonians $H_a$ do not share any variables and can be treated separately. Therefore, we identify $H_a$ as the slices of our parallelized circuit and $H_c$ as the part of the Hamiltonian to only simulate classically within the global Hamiltonian, as we did for $H_{\tilde{C}}$ in \cref{sec: parallelize}.


\section{Numerical results}\label{sec: numerical results}
 To test our algorithms, we solve $50$ randomly generated instances of VRPs with $2$ vehicles and $3$ locations. The locations of the studied instances are generated with a Gaussian distribution over a discrete grid of $100\times 100$. The depot is placed at the center of the grid, with coordinates $(0, 0)$. The distances between locations are computed with the L2 norm. We use NVIDIA's cuQuantum~\cite{cuquantum} to simulate our circuits executed on a DGX-1 with Tesla V100~\cite{dgx}.

Because of the relatively small sizes of VRP instances studied here, we can calculate the global optima with brute force approaches. To evaluate the efficacy of each algorithm we compute the approximation ratios with respect to the brute-force solution. The results obtained by the quantum algorithms are then compared to the open-source software package OR-Tools by Google~\cite{ORTools}.

In addition, to analyze the effect of a lower expressive circuit on the training of the parameters we evaluate the optimized parameters of pQAOA with a QAOA ansatz and we compare the results with the original QAOA performance.

\subsection{Implementation description}
We use two different versions of the pQAOA presented in \cref{subsec: pQAOA}. Each implementation depends on the choice of the number of angles to train in the circuit. The angles $\vec{\gamma}$ and $\vec{\beta}$ can be chosen either to be consistent with QAOA, i.e. each layer is driven by a unique angle, or, since the slices are independent we can use different angles per slice. In the following subsections, we describe the implications of such a choice.

\subsubsection{Multi-angle pQAOA}\label{subsec: multi angle}
 After identifying the slices in the model, each classically separable Hamiltonian is implemented as a separate quantum circuit. The original QAOA has $2$ parameters per layer, but now, since there is no connection between the smaller quantum circuits implemented by the slices, one can decide to assign independent parameters per slice. This yields $2\cdot k\cdot p$ angles to optimize, where $k$ is the number of slices identified in the mode and $p$ is the number of layers of the QAOA circuit. The implementation and the sampling process of the circuit are shown in section \cref{subsec: pQAOA}. 

\subsubsection{Single-slice pQAOA}\label{subsec: single slice}
Differently from what we describe in \cref{subsec: multi angle}, one can decide to not use different parameters per slice but to keep the same number as for the original QAOA. Notice that, when the slices identify identical Hamiltonians this yields a different implementation of the circuit. This happens when the variables used to represent the optimization problem are based on multiple indices. In those cases, the structure of the polynomial $x^T Qx$ presents identical Hamiltonians that are repeated addends in the sums derived by the Cartesian product between the indices. It is worth noting that this is common when discrete variables are implemented employing a binary representation. Indeed, as in the VRP example, one can implement the quantum circuit only considering a single slice. If every slice constructs the same circuit, and the same parameter values are used for each slice, then sampling from the different slices is equivalent to sampling more from one single slice. Therefore, we can implement the circuit of the Hamiltonian that represents the slice and reconstruct the solution of the original Hamiltonian by considering the Cartesian product of the samples with themselves. This means that if the optimization is encoded on a Hamiltonian defined on $N=kn$ qubits, where $k$ is the number of identical slices, we can approximate such Hamiltonian by using only $n$ qubits. Therefore, We call this version of the algorithm, single-slice pQAOA. 

\subsection{Result comparison}
\begin{figure}
    \centering
    \includegraphics[scale=0.5]{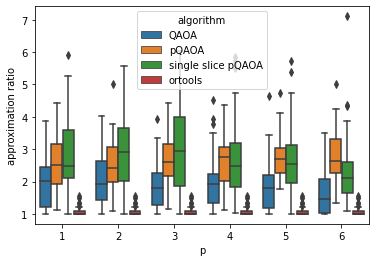}
    \caption{Comparison of the approximation ratios of the algorithms executed on $50$ instances of Gaussian distributed VRP with $2$ vehicles and $3$ locations. To train QAOA we use $10^{p+1}$ per iteration of the classical optimizer while both pQAOA and single-slice pQAOA are optimized by considering only $100$ samples per slice over $10^{p+1}$ samples obtained from the quantum circuit. Therefore the number of samples collected is the same, while the training samples remain fixed only for the pQAOAs. One can notice that while pQAOA achieves better results with smaller $p$, single-slice pQAOA increases the quality of its solution by increasing $p$. Furthermore, we stress the fact  that these small instances are not trivial to solve. In fact, we notice that the classical solver cannot always reach the global optimum of the problem. Nevertheless, all the quantum algorithms perform worse on average.}
    \label{fig:comparison}
\end{figure}

In \cref{fig:comparison} we show the main comparison between the algorithms. Even though the number of samples collected is the same, the number of measurements utilized to produce these samples varies based on the algorithm employed. In the case of QAOA training, we sample from the circuit $10^{p+1}$ times, which is adequate to demonstrate the solution quality trend with the increase of $p$. However, for pQAOAs training, we sample the circuit the same number of times as QAOA and we further subsample from this set to have a fixed number of training samples at each iteration. This decision is made to assess the performance of the algorithms while using fewer resources. Therefore, for multi-angle pQAOA, we sample each slice $10^{p+1}$ times, but we use only $100$ subsamples per slice, which yields a total number of $100^k$ samples to evaluate the global Hamiltonian $H$, where $k$ is the number of slices. In this case, we obtain $10,000$ samples with the Cartesian product because we have two slices. Furthermore, notice that this approach yields the same number of qubit measurements as that of QAOA since the total number of qubits included in the QAOA circuit, in the slices of multi-angle pQAOA, is identical. For single-slice pQAOA, as already mentioned in \cref{subsec: single slice}, we can directly sample the unique slice $10^{p+1}$ times, since the others are identical and, then, reconstruct the solutions to evaluate the global Hamiltonian by considering the Cartesian product between the set of the samples and itself. This yields an advantage because we are using fewer measurements since we do not need to sample from each slice. In addition, by using a unique slice, we are not considering as many qubits as the other algorithms. After the training process, for all algorithms, we sample the circuit to obtain $10,000$ samples to obtain the solution. Therefore, we sample the QAOA circuit $10,000$ times and, to obtain the same number of samples, the two slices of the pQAOAs $100$ times. To summarize, for QAOA, all collected samples are utilized, while for pQAOAs the size of the product between the set of samples from the slices increases exponentially, necessitating the consideration of subsamples. Therefore, since we are considering instances with two slices we consider only $100$ subsamples, out of the $10^{p+1}$ obtained. Notice that the size of a Cartesian product of two sets of size $100$ is a set of size $10,000$. Hence, the amount of resources used to train the parameters is the same for every $p$.

Since the original QAOA has the guarantee to increase the quality of the solutions as the layers of the quantum circuit increase, we execute the algorithms with a different number of layers, $p=1,\ldots,6$, to compare the behavior of the two parallelized algorithms with the QAOA. 

\Cref{fig:comparison} demonstrates that with the classical optimizer reaching convergence for all instances and that despite being given a different amount of training resources, the performance of the pQAOAs is on average worse than QAOA. This was expected because of the smaller number of samples used to train the circuit and the loss of information represented in the circuit. Nevertheless, for some specific instances, we find that the parallelized versions sometimes reach better solutions than QAOA. We can further observe that for small $p$ multi-angle pQAOA performs better than single-slice pQAOA. Nevertheless, the performance of single-slice pQAOA does increase when the circuit becomes deeper and the results surpass the multi-angle pQAOA with $p=6$. Indeed, multi-angle pQAOA does not show any improvement with larger $p$ and its performance stays similar independently from the depth of the circuit. We can attribute this behavior of the algorithm to the number of parameters used to train. We can notice that with lower $p$ a higher number of parameters yields better results, and so multi-angle pQAOA beats its single-slice version. But, on the other hand, having too many parameters leads to a decrease in the quality of the solutions with larger $p$. This observation manifests the NP-hardness~\cite{bittel2021training} of training VQA parameters. Moreover, this is due to the number of samples used to train the parameters, since we use a fixed amount of resources to train. Therefore, the complete distribution of the outcome states of the circuit is not completely described. This issue could be solved either by increasing the number of training samples or by taking into account a state vector representation of the outcome. However, we stress that using a state vector representation is not suitable for real purposes, since we cannot access the final wavefunction of a VQA directly. Indeed, training quantum algorithms considering the wavefunction can be done only in classical simulation and does not have a real-world application.

Indeed, even though our quantum circuit can be now executed by using fewer quantum resources, still the information lost in the process, i.e. the lower expressibility of the model, must be compensated in the classical optimization subroutine. In addition, one can notice that by introducing subsamples, we are also introducing biases in the solutions that we are using to optimize the parameters of the pQAOAs. In fact, since we do not know the original distribution of the outcome of the circuit, we cannot subsample and be sure that the original distribution is preserved. To solve this issue we apply a naive rule to select the subsamples. As already mentioned, the global minima solutions of the global Hamiltonian are not always minima of the slices, therefore we are no longer seeking ground states of the slices. This implies that the parameters must not be trained with the solutions that have a smaller expectation value but rather with solutions that are feasible for the slice. This is because feasible solutions for the global Hamiltonian are also feasible for the slices. Therefore, we only select subsamples of the slices that correspond to feasible solutions for the slice in the global Hamiltonian, including excited states of the slices. If there are no feasible solutions available, we select solutions with the smallest expectation values.

Despite our choice of subsampling, we notice that this classical post-processing represents the main bottleneck of the method. In fact, the ideal training of the parameters requires the use of all the reconstructed solutions via the Cartesian product. This is, though, not practical since the Cartesian product size scales exponentially with the number of slices. Hence, even though we can decide how many slices we want to create with our approach, we must still consider the additional overhead when training with the solutions derived from the Cartesian product. Therefore, better rules to select the subsamples must be found to reduce the biases introduced and improve performance.

\begin{figure}
    \centering
    \includegraphics[scale=0.5]{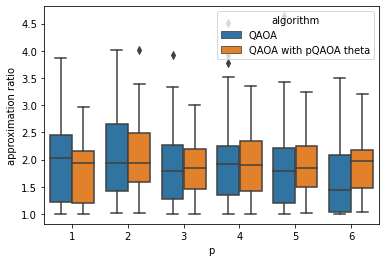}
    \caption{Comparison between QAOA and the results of the QAOA circuit evaluates the best set of parameters trained by using pQAOA. This figure presents the approximation ratio of $50$ instances of the VRP with $2$ vehicles and $3$ locations. The results of QAOA are the ones shown in \cref{fig:comparison}. To generate the solution of QAOA with the parameters of pQAOA we evaluate a QAOA circuit evaluates on the parameters of each slice and we pick the best results. We notice that the results are similar for lower depths of the circuit.}
    \label{fig:pqaoa theta}
\end{figure}

\begin{figure}
    \centering
    \includegraphics[scale=0.5]{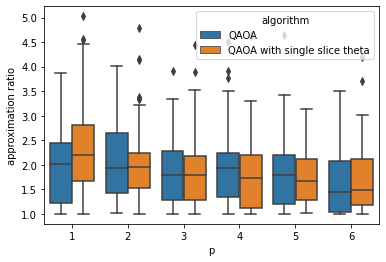}
    \caption{The figure presents a comparison similar to \cref{fig:pqaoa theta}, but now we evaluate the QAOA circuit with the parameters trained by the single-slice pQAOA. We can see that the results match the performance of QAOA for deeper circuits. Furthermore, we stress the fact that the quality of the solutions increases by increasing the depth of the circuit.}
    \label{fig:single slice theta}
\end{figure}

The presented results can be used to further analyze QAOA. In \cref{fig:pqaoa theta} and \cref{fig:single slice theta} we compare the results between QAOA and the same QAOA circuit evaluated with sets of parameters trained by pQAOA. The approximation ratio of the quantum algorithms is comparable. We attribute this similarity to the concentration of parameter phenomenon~\cite{brandao2018fixed}. As already highlighted, we can notice that the results obtained by using multi-angle pQAOA have higher quality solutions with small $p$ while single-slice pQAOA obtains better results for deeper circuits. Furthermore, the quality of the solutions improves by increasing the number of layers implemented in the circuit. The same trend is shown by the results computed by evaluating a QAOA ansatz with trained parameters obtained from multi-angle pQAOA and single-slice pQAOA.

Additionally, it is worth noting that the performance of all the quantum algorithms is worse than the classical results. Moreover, the instances appear not to be trivial since the classical solver does not always reach the optimal solution. Nevertheless, we stress that in the considered examples the size of the instances can be considered small and, therefore, an advantage is not expected. However, we put the reference to standard methods in classical optimization. Lastly, it is important to note that the optimal solutions of all the instances require only one vehicle to leave the depot. In fact, albeit generating Gaussian distributed instances is standard while benchmarking VRP instances, with small instances we experience this bias. This is due to the rareness of selecting locations on the grid that are far enough to allow both vehicles to leave the depot. Indeed, sampling from the tails of the Gaussian distribution is difficult when the number of samples is small, as is the case for our instances.


\section{Conclusions}





In this work, we present a method to create parallelized versions of quantum algorithms informed by the optimization problem directly. Our analyses were focused on one specific algorithm, QAOA, but they can hold for all the VQAs presented in this work. We show how to construct parallel quantum circuits to maximally utilize the available number of qubits in NISQ processors. Specifically, we show how to use this method to solve constrained optimization problems that have more variables than the number of qubits available in the QPU. Furthermore, in specific classes of optimization problems, especially in constrained optimization problems, each parallel slice obtained by this process creates an identical copy of the same quantum circuit, which we call single-slice pQAOA. We show how to even further reduce the need for quantum resources by simulating only one of the identical copies.

We find that for low-depth circuits (specifically for $p=1,2$) our parallelization method of multi-angle pQAOA is comparable with QAOA and in some cases even better. In fact, even though the circuit is less expressive than the QAOA one, the larger number of parameters returns better results from the optimization routine. On the other hand, when the depth increases we see that a large number of parameters becomes a bottleneck due to the hardness of finding optimal values. This is stressed by the performance of the single-slice pQAOA as well, that for larger $p$ becomes, instead, competitive with QAOA performance. Indeed, we can notice that while the approximation ratio for small $p$ does not show any improvement with respect to multi-angle pQAOA, we can see that this changes with deeper circuits. Therefore, we can highlight a trade-off between the number of parameters to optimize and the depth of the circuit that represents the model.

In addition, we can see that the number of qubit measurements that we use to sample the circuits varies. For QAOA and multi-angle pQAOA we use the same while for single-slice pQAOA we need polynomially fewer measures. Therefore this scaling makes single-slice pQAOA a good candidate to make quantum hardware a valuable alternative to solve real-world problems since it is more practical to implement problems at scale.

It is also worth highlighting that while a more expressive model can yield better results, training the QAOA circuit over the entire model may not be necessary. Specifically, a less expressive quantum circuit with a reduced number of gates can produce parameters that yield solutions comparable to those obtained by training over the original model. Notably, the quantum resources required to obtain this set of parameters are lower than those needed for the original QAOA. The number of qubits and measurements required to compute the parameters is also lower than that of the non-parallelized algorithm. These results have significant implications for the design and optimization of quantum circuits for practical applications.

Therefore, scaling the problem by reducing or keeping the amount of quantum resources while obtaining comparable results makes this method a possible solution for the application of quantum algorithms to solve real-world problems. Furthermore, future research should focus on developing new classical methods for reconstructing the original global Hamiltonian and developing tailored rules for properly collecting subsamples to achieve a scalable implementation of VQAs.


\section*{Acknowledgements}
MC and SY are funded by the German Ministry for Education and Research(BMB+F) in the project QAI2-Q-KIS under grant 13N15587. Furthermore, the authors would like to thank  Andrea Skolik, Anestis Papanikolaou, Gabriele Compostella, Jakob Huhn and Matthew Kiser for valuable discussions and suggestions
given.

\bibliography{lib}

\appendix

\section{Multi-objective QAOA}
Though \cref{sec: parallelize} is mainly focused on analyzing the optimization problems to find slices by looking at the constraints, we want to stress that this identification can be done by looking at other structures in the optimization problem. Indeed, by exploiting the method presented in \cite{hao2022exploiting} we can choose to include in $H_{\tilde{C}}$ also the objective function of the problem inspected. This can lead to a further decomposition of the problem. For instance, if we consider the VRP, we can see that if we consider $H_{\tilde{C}} = H_\mathrm{of} + H_{c}$ to be the term to only simulate classically, we have that the slices are now the Hamiltonians
\begin{equation*}
    H_a = \sum_{s=0}^{n}\left(1-\sum_{i=0}^{n} x_{a,i,s} \right)^{2}.
\end{equation*}
Therefore, the quantum circuits only implement the one-hot constraint, which ensures that a vehicle can only be in one location at a given time step.

Nevertheless, considering the objective function as part of $H_{\tilde{C}}$ allows us to implement multi-objective constraint optimization problems. Let us consider the following mathematical program in the standard form:
\begin{mini*}
    {x\in\mathbb{B}^{n}}{(f_i(x))_{i=1,\ldots, m}}{}{}
    \addConstraint{\sum_{j=1}^{n}c_{ij} x_{j}}{=b_i\quad}{\forall i\in[k].}
\end{mini*}
Let us now formalize the Hamiltonian that only implements the constraints. This is the slice in this example and it is the only one that we identify. The Hamiltonian can be read as:
\begin{equation}\label{eq: contraint ham}
    H = \sum_{i=0}^{k}\left(\sum_{j=1}^{n}c_{ij}x_j - b_i \right)^{2}.
\end{equation}
We can notice that, by following the same procedure as in \cref{fig: pQAOA}, we can implement a QAOA circuit that represents the Hamiltonian $H$ and we optimize the circuit over the function:
\begin{equation*}
    \left(f_i(x) + H \right)_{i=1,\ldots, m}.
\end{equation*}
Since now we are evaluating a multi-objective function, we have to evaluate and identify the parameter based on the Pareto front. After the selection of the "optimal" point, the parameters are evaluated and the quantum circuit is updated. This allows the implementation of multi-objective combinatorial optimization problems without using a metric to reduce the multi-objective function to a QUBO problem.

\section{Slicing with different rules}\label{sec: non constraints}
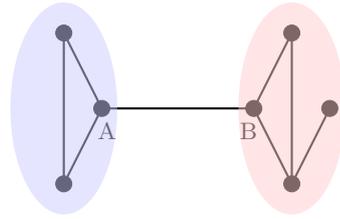
\begin{figure}
    \centering
    \begin{tikzpicture}[scale=2.0]
    
    \filldraw (-0.5,0) circle(1.5pt) node[right=2pt, below=2pt] {A};
    \filldraw (-0.75,0.5) circle(1.5pt);
    \filldraw (-0.75,-0.5) circle(1.5pt);
    
    \filldraw (0.5,0) circle(1.5pt) node[left=2pt, below=2pt] {B};    
    \filldraw (0.75,0.5) circle(1.5pt);
    \filldraw (0.75,-0.5) circle(1.5pt);
    \filldraw (1.0,0) circle(1.5pt);

    \draw[thick] (-0.5,0)--(0.5,0);
    
    \draw[thick] (-0.5,0)--(-0.75,0.5);
    \draw[thick] (-0.5,0)--(-0.75,-0.5);
    \draw[thick] (-0.75,0.5)--(-0.75,-0.5);
    \fill[color=blue!20, opacity=0.5] (-0.75, 0) ellipse [x radius= 10pt, y radius=20pt]; 

    \draw[thick] (0.5,0)--(0.75,0.5);
    \draw[thick] (0.5,0)--(0.75,-0.5);
    \draw[thick] (0.75,0.5)--(0.75,-0.5);
    \draw[thick] (1.0,0)--(0.75,-0.5);
    \fill[color=red!20, opacity=0.5] (0.75, 0) ellipse [x radius= 10pt, y radius=20pt];
    
    \end{tikzpicture}
    \caption{We consider the MaxCut problem over the considered graph. In the picture, we highlight the two slices that we build in this example. The label $A$ and $B$ are use to identify in the Hamiltonian $H$ the term $H_{\tilde{C}}$.}
    \label{fig: MaxCut}
\end{figure}

As mentioned in \cref{sec: parallelize} the slicing method can be applied by considering any structure informed by the optimization problem, not only constraints or the objective function. Let us show this by considering an example. We consider the MaxCut problem on the graph shown in \cref{fig: MaxCut}. The Hamiltonian of the MaxCut problem~\cite{farhi2014quantum} is define as
\begin{equation*}
    H=-\sum_{(i,j)\in E}s_is_j,
\end{equation*}
where $E$ is the set of edges of the graph considered and $s_i$ is the spin variables that take value $+1$ if the node $i$ is considered in one partition of the MaxCut and $-1$ otherwise. 

In \cref{fig: MaxCut} we can identify two different regions of the graph that are connected only by one edge. In this case, we can read the Hamiltonian in the following fashion:
\begin{equation*}
    H= -s_As_B + \left(-\sum_{(i,j)\in E_b}s_is_j - \sum_{(i,j)\in E_r}s_is_j\right),
\end{equation*}
where $E_b$ is the set of edges identified by the blue region and $E_r$ is the set of edges identified by the red region. By using the same notation as \cref{sec: parallelize}, we can easily see that $H_{\tilde{C}}=-s_As_B$ and the two addends inside the brackets are the two slices. In this case, we can apply the slicing method because of a topological property of the graph. In fact, we can observe that the graph is $1$-connected, i.e. the graph becomes disconnected by leaving one edge out. As we already notice, one can identify the edge $(A,B)$ as the edge that connects the two connected components defined by the edge sets $E_b$ and $E_r$.

It is important to notice that this results in the same circuit cutting method for QAOA presented in \cite{marshall2022high}. The only difference is in the choice of the gates to leave out. In our method, the choice is informed by the problem, while in \cite{marshall2022high} the optimal cut must be found introducing a problem that reduces to the graph partitioning problem, which is NP-hard~\cite{graphpartitioning}. 

The MaxCut problem can be considered a special example because the quadratic terms defined in the Hamiltonian, as well as the two-qubit gates of the QAOA ansatz, has a one-to-one connection with the edges of the problem graph. Therefore the connected subgraphs correspond exactly to the slices of the circuit as well. However, in general, this example shows that we can exploit different properties of the graph defined by the optimization problem (e.g. the $k$-connectivity, as in this example) in order to apply the slicing method.

\end{document}